\documentclass[12pt]{iopart}
\usepackage{graphicx}
\begin{document}

\title[Resonant light interaction with plasmonic nanowire systems]{Resonant light interaction with plasmonic nanowire systems} 

\author{Viktor A. Podolskiy$^{1}$
\footnote[3]{To whom correspondence should be addressed (vpodolsk@princeton.edu)}, 
Andrey K. Sarychev$^{2}$, 
Evgenii E. Narimanov$^{1}$
Vladimir M. Shalaev$^{3}$}

\address{$^{1}$ Electrical Engineering Dept., Princeton University, Princeton, NJ 08540}

\address{$^{2}$ Ethertronics Inc., 9605 Scranton Road, Suite 850
San Diego, CA 92121}

\address{$^{3}$ School of Electrical and Computer Engineering, Purdue University, West Lafayette, IN 47907}

\begin{abstract}
We compare the optical response of isolated nanowires, double-wire systems, and $\Pi$-structures, and show that their radiation is well described in terms of their electric and magnetic dipole moments. We also show that both dielectric permittivity and magnetic permeability can be negative at optical and near infrared frequencies, and demonstrate the connection between the geometry of the system and its resonance characteristics. We conclude that plasmonic nanowires can be employed for developing novel negative-index materials. Finally, we demonstrate that it is possible to construct a nanowire-based ``transparent nanoresonator'' with dramatically enhanced intensity and metal concentration below 5\%. 
\end{abstract}

\maketitle

\section{Introduction}
The concept of light manipulation on the subwavelength scale is increasingly attractive to researchers in optics, materials science, and chemistry\cite{shalaev,sibiliaBook,others}. Plasmonic materials, strongly interacting with light are the exceptional candidates for the nano-photonic devices. It has been shown that metal-dielectric films can be used to effectively confine the electromagnetic wave to a nanoscale, and transmit the localized signal~\cite{percolation,shalaev,podolskiyHoles}. Another class of nanoplasmonic devices based on metallic nanowires and nanowire pairs, have been suggested to obtain light nano-confinement, transmission, and even negative refraction for optical and infrared frequencies~\cite{PodolskiyJNOPM,PodolskiyOptExp2003} due to resonance excitation of electric and magnetic dipole moments. However, while the dipole contribution dominate the forward and backward scattering of the nanowire systems, the presence of quadrupole moment may substantially affect their ``side'' scattering characteristics. 

Here we address the electromagnetic response of the nanowire systems, and show that it is indeed dominated by the dipole moments. We also demonstrate that the electric and magnetic resonances correspond to a special kind of surface (plasmon polariton) wave, and can be independently controlled using the combination of nanowires and $\Pi$-structures\cite{Sarychev2004}. Finally, we show that the nanowire composites may be used to build a ``broadband transparent nanoresonator'', achieving an {\it average} intensity enhancement exceeding an order of magnitude, with a metal concentration less than $5\%$. 
The rest of the paper is organized as follows: in we first briefly describe the typical nanowire geometry and the coupled dipole equations (CDEs), used in our simulations. We then describe the resonant response of a single nanowire, nanowire pair and $\Pi$-structures. Finally, we show the field enhancement in the {\it transparent} nanowire composite. 

\section{Simulation of the nanowire response using CDEs} 

\begin{figure}
\centerline{\includegraphics[width=6.5cm]{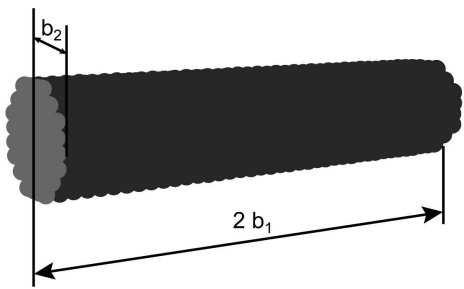}}
\caption{\label{figWire} A long nanowire represented by the array of point dipoles.}
\end{figure}

The typical radius of individual nanowires described in this manuscript $b_2$ is much smaller than the wavelength of the incident light $\lambda$, and is comparable with the skin-depth of the material. The length of the wire $2 b_1$, on the other hand, is comparable to the wavelength (see Fig.\ref{figWire}). The electromagnetic properties of such metallic nanostructures are somewhat similar to the properties of scaled-down radio-antennas widely used in telecommunications. However, the finite value of the dielectric constant of the metal in the optical range and typically low aspect ratio lead to the fundamental differences in the response of optical- and radio- antennas. These differences make the analytical solution for the problem of electromagnetic response of nanowires hardly possible. Here we use the well-known coupled-dipole approximation\cite{Purcell,draine,PodolskiyOptExp2003} to find the response of our system. 

In this approach, a single nanowire is represented by an array of point dipoles arranged at the sites of a cubic lattice. Each dipole is subjected to the field of an incident plane wave and to the field of all other dipoles. Thus, the dipole moments of all dipoles are coupled through the following {\it coupled-dipole equations} (CDEs): 
\begin{equation}
\label{eqCDE}
{\bf d}_i=\alpha_0 \left[
{\bf E}_{inc}+\sum_{j\neq i}^N \hat{G}({\bf r}_i-{\bf r}_j){\bf d}_j
\right], 
\end{equation}
where $E_{inc}$ represents the incident field at the location of $i$-th dipole, ${\bf r}_i$, $\hat{G}({\bf r}_i-{\bf r}_j){\bf d}_j$ represents the EM field scattered by the dipole $j$ at this point, and $\hat{G}$ is a regular part of the free-space dyadic Green function defined as
\begin{eqnarray}
G_{\alpha\beta}=k^3[A(kr)\delta_{\alpha\beta}+B(kr)r_{\alpha}r_{\beta}],
\nonumber\\
\label{eqGreen}
A(x)=[x^{-1}+ix^{-2}-x^{-3}]\exp(ix),\\
B(x)=[-x^{-1}-3ix^{-2}+3x^{-3}]\exp(ix),\nonumber
\end{eqnarray}
with $\hat{G}{\bf d}=G_{\alpha\beta}d_{\beta}$. The Greek indices represent the Cartesian components of vectors and the summation over the repeated indices is implied. 

The key parameter in CDEs is the polarizability of a monomer, $\alpha_0$, usually given by Clausius-Mossotti relation (see, e.g.~\cite{Jackson}) with the radiative correction introduced by Draine \cite{draine}
\begin{eqnarray}
\alpha_{LL}=R^3\frac{\epsilon-1}{\epsilon+2},\\
\alpha_0=\frac{\alpha_{LL}}{1-i(2k^3/3)\alpha_{LL}},
\end{eqnarray}
where $\epsilon$ is the dielectric permittivity of the material and $\alpha_{LL}$ is the Lorentz-Lorenz polarizability without the radiation correction. The magnitude of this polarizability, controlled by the parameter $R$ serves as a fitting parameter. It may be visualized as a radius of an imaginary sphere, centered over the position of a point dipole. In our simulations this parameter is determined by the condition that the system response in the quasi-static limit should yield the correct depolarization factors, and is typically varied in the range $R=0.59 \;...\; 0.62$ (in the units of the lattice size). 

\section{Radiation and resonance properties of metallic nanowires}

The non-resonant scattering by a nanostructured system typically weakly affects the incident electromagnetic radiation, and can be effectively treated by a variety of mostly perturbative techniques (see, e.g. Ref.~\cite{Jackson}). However, when the frequency of incident light coincides with a resonant frequency of a nanostructure, the electromagnetic field distribution may be dominated by the scattered (radiated), and not the incident wave. The scattered by a nanostructure field can be expanded into the series of multipole components. First term in such an expansion corresponds to electric dipole. Next two terms correspond to magnetic dipole and electric quadrupole~\cite{Jackson,PodolskiyOptExp2003}. 

\begin{figure}
\centerline{\includegraphics[width=12cm]{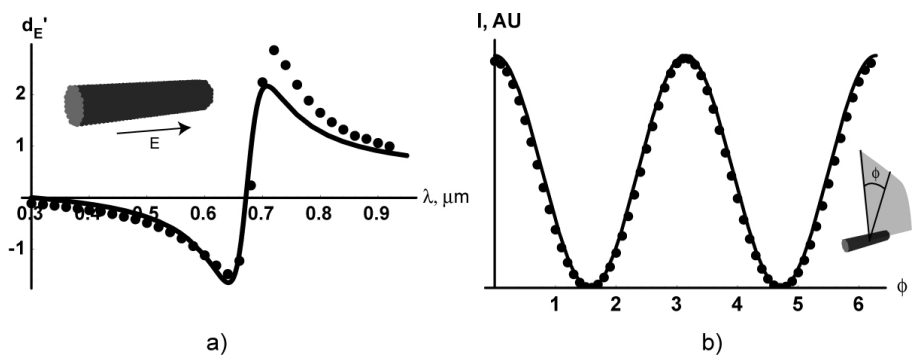}} 
\caption{\label{figDipole} (a) The comparison between the dipole moment found from numerical simulations (dots) and calculated using Eq.\ref{eqMom} (line). The moments are normalized by unit volume. (b) Far-field intensity radiation pattern of a single nanowire in (a) is obtained from numerical simulations (dots) and calculated by approximating the antenna by a point dipole (solid line). The dimensions of the silver wire are $162\times 32 \times 32$nm. The far field pattern is calculated for $\lambda=560$nm. }
\end{figure}

Similarly to the well-known case of radio-antennas~\cite{Jackson}, the radiation properties of an isolated nanowire are well-described by its dipole moment (Fig.~\ref{figDipole}). The induced polarization in a substantially long and thin wire close to its first resonance ($2b_1=\lambda_p/2$, where $\lambda_p$ is the wavelength of the plasmon polariton) can be represented by the following relation \cite{Sarychev1996}:
\begin{eqnarray}
d_E=\frac{2}{3}\left[b_1b_2^2f(\Delta)E\epsilon_m \right]/ 
\label{eqMom}
\left[1+f(\Delta)\epsilon_m\frac{b_1^2}{b_2^2} \ln\left(1+\frac{b_1}{b_2}\right)\cos\Omega\right], 
\end{eqnarray}
where the dimensionless frequency $\Omega$ is given by $\Omega^2=(b_1k)^2\frac{\ln(b_1/b_2)+ikb_1}{\ln(1+b_1/b_2)}$. The function $f(\Delta)=\frac{1-i}{\Delta} \frac{J_1[(1+i)\Delta]}{J_0[(1+i)\Delta]}$ is introduced to account for the skin effect; the parameter $\Delta=b_2\sqrt{2\pi\sigma_m\omega}/c$ represents the ratio of nanowire radius and the skin depth, and $\sigma_m$ is the bulk metal conductivity. 

The resonances in an isolated nanowire can be related to a resonant excitation of a special kind of surface waves on the metal -- air interface. These waves, which exponentially decay away from the interface are also known as plasmon polaritons. Although the excitation of the plasmon polariton with a plane electromagnetic wave is impossible in the infinite medium (see e.g.~\cite{landauECM}), the plasmon waves have non-zero resonance width in {\it finite} nanowires so they can effectively couple to a plane wave. The frequency of the polariton resonance is controlled by the nanowire material and length, while the ``width'' of this resonance is related to the skin-depth and the radius of the wire~\cite{PodolskiyOptExp2003}. This effect makes it possible to use a single metallic nanoantenna to confine and transmit the optical and infrared radiation {\it on the nanoscale}~\cite{PodolskiyJNOPM,PodolskiyOptExp2003}. 

\begin{figure}
\centerline{\includegraphics[width=12cm]{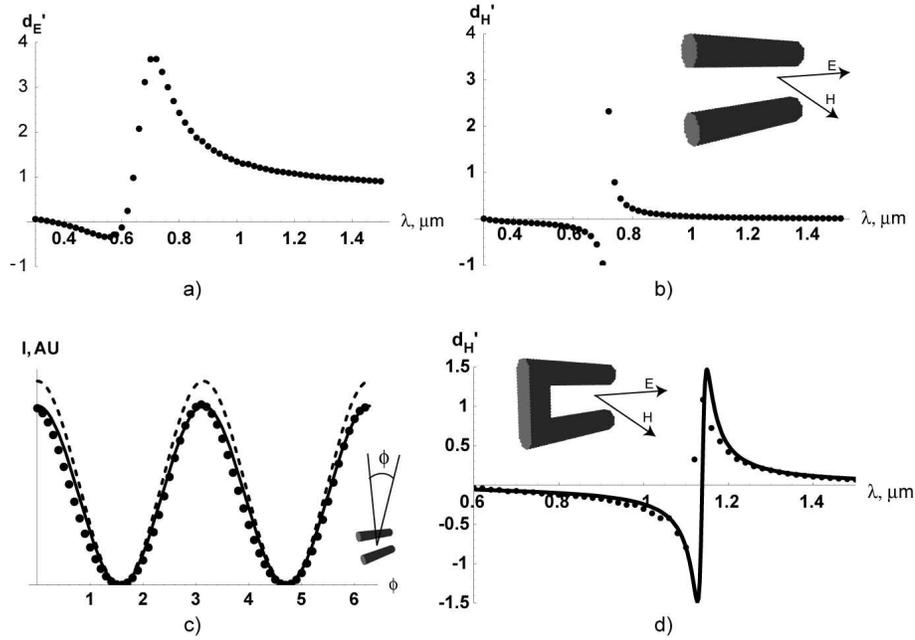}}
\caption{\label{figPair} (a) The dipole moment of the coupled nanowire system with dimensions 162 nm (antenna length) by 32 nm (antenna diameter) by 80 nm (distance between antennas). (b) The magnetic dipole moment of the system in (a). (c) Far field intensity distribution of system in (a-b) calculated using exact numerical simulations (dots), by approximating the system by point electric dipole (dashed line) and by approximating the system by point electric and magnetic dipoles (solid line). The far field pattern is calculated for $\lambda=560$nm. (d) Connecting the coupled nanowires in (a-b) into a $\Pi$-structure drastically shifts the position of the magnetic resonance, leaving the dipole moment of the system practically unchanged (not shown). The dots correspond to numerical simulations, solid line corresponds to Eq.\ref{eqPi}. All moments are normalized by the unit volume. }
\end{figure}

The behavior of the plasmon modes changes substantially when two nanowires positioned closely to each other, so that their plasmon modes can interact with each other. When the electric field of the incident plane wave is parallel to the wires, and the magnetic field to be perpendicular to the common plane of the two wires (see Fig.\ref{figPair}), two kinds of plasmon polariton waves of the different symmetry can be excited. 

The symmetric combination of the two polariton waves leads to the excitation of the dipole moment in both wires. In this case, the electric field of the incident plane wave resonantly excites parallel currents in both nanowires. The anti-symmetric combination, on the other hand, corresponds to the {\it anti-parallel currents} in the two wires (excited by a magnetic field component of the incident wave). These currents, together with the displacement currents in between the wires, lead to a resonance {\it magnetic moment} in the system. The shift between the resonance frequencies of electric and magnetic dipoles is related to a coupling efficiency between the polariton modes in two wires, which in turn is controlled by the distance between the wires. This effect has a similar nature to the splitting of the energy levels of the wavefunctions of different symmetry in the double-well potential due to the tunneling coupling. 

The effect of the polariton modes interaction on resonance characteristics is clearly seen when we compare the electric and magnetic response of the system of two parallel nanowires and of $\Pi$-system of the same size, which is obtained by bringing one end of the two wires into the electric contact [see Fig.~\ref{figPair}(d)]. As it is explained above, the electric resonant response of the system is governed by symmetric polariton mode. The currents in the two nanowires in this case have essentially the same distribution, so they are not affected by the electric contact between wires. The magnetic response of the two systems, on the other hand, is dramatically different, since the presence of the electric contact forces the connected points to have the same value of the potential, and makes it the excitation of anti-symmetric polariton mode impossible. 

Due to the presence of the electric contact, the magnetic resonances of the $\Pi$-structure do not directly correspond to the electric resonances of the single nanowire. It can be shown that $\Pi$-structure may have resonant magnetic response even when its size is much smaller then the wavelength, so that no polariton modes can propagate on the corresponding nanowire. The magnetic resonance in this case is similar to electric plasmon resonance which occurs in all metallic nanoparticles. In the limit of $\lambda\gg b_1 \gg d \gg b_2 $ the magnetic resonance of the $\Pi$ structure can be described by the following expression~\cite{Sarychev2004}
\begin{eqnarray}
\label{eqPi}
d_H=\frac{1}{2}H_0 b_1^3 \log(d/b_1) (k d)^2 \frac{\tan(gb_1)-gb_1}{(gb_1)^3},
\end{eqnarray} 
where $g^2\approx-4\log(d/b_2)/(b_2^2/\epsilon)$. Note that this resonance frequency is defined solely by the geometry of the system. 

As a result of two-wire interaction, the scattering (far-field) response of the coupled wire systems and $\Pi$-structures, which defines the interaction between different elements in the macroscopic composite, is not described by a dipole moment alone. As it is shown in Fig.~\ref{figPair}, such systems have magnetic dipole moment comparable to their electric dipole moment. Our numerical simulations clearly show that these systems are fully described by two dipole moments, and have vanishing higher moments. This in turn leads to highly-directional emission (scattering) properties of the double-wire systems. This fact is illustrated in Fig.\ref{figPair}, which shows excellent agreement between the far field obtained by numerical calculation and by approximating the system by point electric and magnetic dipoles. 

The electric and magnetic resonances can be used to produce extremely large local fields, which may be beneficial for a variety of spectroscopic, lithographic, and biological applications as described in the next chapter. 

However, one of the most promising applications of nanowire composites lies in the area of materials with simultaneously negative dielectric permittivity and magnetic permeability. Such media, originally considered by Veselago \cite{veselago} was predicted to have a negative refractive index and consequently exhibit a wide variety of surprising optical phenomena. Among them are the reversed Snell's law~\cite{LHM}, Cherenkov radiation, and Doppler Effect. Due to its negative phase velocity such media are often referred to as ``left-handed'', meaning that the wavevector and the vectors of electric and magnetic fields form in such a material a left-handed trio in contrast to a conventional ``right-handed'' case. One of the most promising phenomenon present in left-handed materials is so-called ``superlensing'', where a slab of a medium with $\epsilon=\mu=-1$ is used to obtain an optically perfect image with subwavelength resolution in the far field \cite{pendry2000,imaging}. 

As we showed above, the system of coupled nanoantennas exhibit resonance electric and magnetic response. When the wavelength of an incident light is below the resonance in the coupled nanowire system, the excited moments are directed in opposite to the excitation field (Fig.\ref{figPair}). Such negative responses may be used to implement a left-handed composite in the optical and near-infrared ranges\cite{PodolskiyOptExp2003,PodolskiyJNOPM,SarychevSPIE,Sarychev2004}.

By changing the geometry of the system we can shift the resonances to any region from the visible to near infrared frequency ranges \cite{PodolskiyOptExp2003,PodolskiyJNOPM,SarychevSPIE}. 

\section{Enhanced local fields in nanoantenna arrays}

As it was mentioned above, the resonance coupling between the plane and polariton waves in a single metallic nanowire opens a possibility to propagate the optical light through the array of nano-antennas, using them as wires in all-optical computers and telecommunication systems. It also opens the possibility of resonant light amplification on the nanoscale due to resonant excitation of polariton waves. 

\begin{figure*}
\centerline{\includegraphics[width=13.5cm]{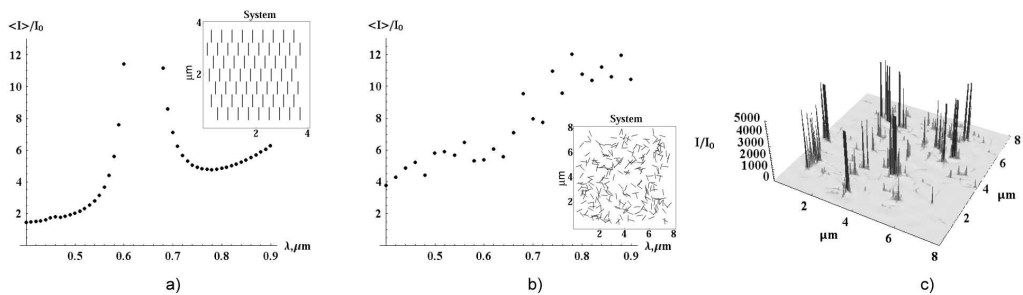}}
\caption{\label{figFields} (a) The average intensity enhancement over the parallel wire composite (inset) clearly shows a separated-resonance structure, an implicit property of single nanowire. (b) The random nanowire percolation composite (inset) exhibits a broadband intensity enhancement due to collective excitation of a large number of different resonant clusters. The intensity distribution over this composite for $\lambda=800\;nm$ is shown in (c). The size of individual wire in both composites is $600\times 20\times 20$nm; surface concentration of metal is 4\%.}
\end{figure*}

The local intensity in this case could exceed the intensity of the incident field by three or more orders of magnitude (Fig.~\ref{figFields}). Such high local fields are beneficial for enhanced spectroscopy, lithography, absorption, nonlinear processes and all related applications. Note that high local fields at the resonance are usually accompanied by the narrow frequency band where the resonance exists \cite{PodolskiyOptExp2003}. The resonant frequency itself is controlled by the length of the nanowire and its material.

While the area of the enhanced local field is concentrated near the nanoantenna surface, the collective resonance of several antennas can lead to the enhancement of the {\it average} intensity in the antenna composite. The response of the equally separated parallel nanowires resembles the behavior of an isolated antenna (Fig.~\ref{figFields}~a), exhibiting huge intensity enhancement in the narrow frequency ranges corresponding to the ``eigen'' frequencies of the plasmon polariton waves in the individual wires. 

The situation changes dramatically when the antennas are randomly deposited on a dielectric substrate, and a surface metal concentration reaches the value when the composite starts to conduct a DC current (known as the percolation threshold). At this point the composite contains nanowire clusters of all possible sizes and configurations, each having its own resonance frequency, and generating at this frequency high local fields. The collective effect of all the clusters leads to the extremely broadband intensity enhancement, as shown in Fig.~\ref{figFields}~(b). This effect is similar to the broadband field excitation in a conventional percolation film \cite{percolation,shalaev}. However, in contrast to the percolation film, where the percolation threshold concentration is fixed and is equal to $50\%$, the percolation threshold in a nanowire composite is inversely proportional to the aspect ratio of the individual nanowires and can be made arbitrary small, making it possible to fabricate a ``transparent nanoresonator''. This system can be effectively used in the areas which require simultaneously high fields and optical transparency, e.g. in stacked solar cells~\cite{Forrest}, transparent bio-sensors or optical lithography. 

\section{Conclusion}
The unique resonant characteristics of metallic nanoantennas could be precisely controlled by their geometry and material properties. The polariton resonance frequency in such devices can be tuned to any given range from the optical to the mid-infrared. Applications of plasmonic nanowire composites include narrow- and broadband nano-resonators, photonics, and left-handed media. 

This work was supported in part by NSF under ECS-0210445, DMR-0134736, and DMR-0121814 grants

\end{document}